\documentstyle[prl,multicol,aps,axodraw,epsf]{revtex}

\newcommand{\be}{\begin{equation}}
\newcommand{\ee}{\end{equation}}
\newcommand{\ba}{\begin{eqnarray}}
\newcommand{\ea}{\end{eqnarray}}
\newcommand{\tr}{{\rm Tr\,}}
\newcommand{\ii}{{\rm i}}
\newcommand{\trq}{{\rm \hat{T}r\,}}
\newcommand{\ex}{{\rm e}}
\newcommand{\nn}{\nonumber}

\newcommand{\bfp}{{\bf p}}
\newcommand{\bfx}{{\bf x}}
\newcommand{\bfy}{{\bf y}}
\newcommand{\bfz}{{\bf z}}

\newcommand{\h}{\cal{H}}

\newcommand{\hp}{\hat{P}}
\newcommand{\htm}{\hat{T}}
\newcommand{\op}{{\cal O}}
\newcommand{\eq}{Eq.~}
\newcommand{\eqs}{Eqs.~}
\newcommand{\fig}{Fig.~}

\begin{document}

\draft        

\title{Non-perturbative singularity structure of gluon and quark propagators}
\author{Owe Philipsen}
\address{
Center for Theoretical Physics, Massachusetts Institute of Technology,\\
Cambridge, MA 02139-4307, USA\\
email: philipse@mitlns.mit.edu}


\maketitle

\begin{abstract}
A gauge invariant, non-local observable is constructed in lattice pure gauge theory, 
which is identical to the gluon propagator in a particular gauge. The transfer matrix
formalism is used to show that this correlator decays exponentially with 
eigenvalues of the Hamiltonian. 
This implies a gauge invariant singularity structure of the propagator
in momentum space, permitting a non-perturbative definition of a parton mass.
The relation to gauge fixing and the extension
to matter fields are discussed.
\end{abstract}

\pacs{PACS: 11.15, 11.15.Ha, 12.38.Aw, 12.38.Gc  \hfill MIT-CTP-3154}

\begin{multicols}{2}

The confinement problem of QCD consists of the dynamical relation between perturbative
parton physics at short distances and non-perturbative hadron physics at large distances. 
To understand
the transition of a system from one regime to the other requires a non-perturbative
study of the dynamics of colour degrees of freedom, which is encoded in the Green functions
of quark and gluon fields, the fundamental degrees of freedom of the action.
This is particularly essential in the context of finite temperature
and density physics probed in experimental heavy ion collisions, where creation
of a ``deconfined'' quark gluon plasma is expected, whose collective physical properties should 
be determined by parton dynamics.

In perturbation theory one fixes a gauge and studies parton interactions directly. 
Because of their gauge dependence, field propagators are not physical observables. 
Nevertheless, physical 
information about the parton dynamics is carried by their singularity structure. 
For example,
the pole mass defined from the quark propagator is gauge
independent and infrared finite to every finite order in perturbation theory \cite{kro}. 
A similar result holds for the gluon propagator, provided an appropriate resummation
of infrared sensitive diagrams has been performed \cite{kkr}. Gauge invariant resummation 
schemes have been designed to self-consistently compute the pole of the gluon propagator
in three dimensions \cite{mm}, which is related to the ``magnetic mass'' regulating
the non-abelian thermal infrared problem \cite{ir}. 
In a Hamiltonian
analysis of the three dimensional gauge theory a gauge invariant composite gluon
variable has been constructed, which in the weak and strong coupling limits yields
a gluon mass gap as the lowest eigenvalue of the Hamiltonian \cite{nair}. 

However, perturbation theory and resummation methods are limited by the requirement
of weak coupling, and nothing is known about the non-perturbative existence
of field propagator poles in a confining regime.
Hence a non-perturbative analysis 
in the framework of lattice gauge theory is warranted. Unfortunately, 
numerical gauge fixing on the lattice \cite{mo} is plagued by several problems:
First, it is difficult to fix a gauge uniquely and avoid
the problem of Gribov copies \cite{grib}. 
Second, most complete gauge fixings (e.g. the Landau gauge)  
violate the positivity of the transfer matrix, thus obstructing
a quantum mechanical interpretation of the results.
Because of these problems, many results from gauge fixed simulations 
have remained controversial. 
An overview with references to recent numerical work
may be found in \cite{rev}. 

For the extraction of quark masses gauge fixing can
be circumvented by the methods of non-perturbative renormalization, which make use
of manifestly gauge invariant quantities \cite{npr}.
In this approach quark masses are defined through
PCAC relations rather than by a field correlator,
but this does not permit to address the same question
for gluons. 

In this letter, the transfer matrix formalism \cite{cre,pos} is used to show 
that field correlation functions 
carry gauge invariant information non-perturbatively.
To this end, $SU(N)$ pure gauge theory with Wilson action 
is considered in $d+1$ dimensions on
a $(aL)^d\times aN_t$ lattice with periodic boundary conditions.


A gauge invariant composite and local field carrying the quantum numbers of a gluon can be 
constructed in theories with a complex scalar $N$-plet in the fundamental representation,
such as the electroweak sector of the standard model.
In the pure gauge theory no such field is present. However, it is possible to construct
complex functions  of the gauge field $f_\alpha[U], (\alpha=1,\ldots N)$, with the same 
transformation behaviour from the covariant Laplacian, 
\be \label{lev}
-\left(D_\mu^2[U]\right)_{\alpha\beta}f^{(n)}_\beta(x)=
\lambda_n f^{(n)}_\alpha(x),  \quad \lambda^n>0.
\ee
The latter is a 
hermitian operator with a strictly positive spectrum, 
whose eigenvectors have the desired transformation property $f^{(n)g}(x)=g(x)f^{(n)}(x)$.
They provide a unique mapping $U\rightarrow f[U]$ except when eigenvalues are degenerate
or $|f|=0$. In practical simulations the probability of generating such 
configurations is essentially zero \cite{vi}. 
These properties have been used previously for gauge fixing 
without Gribov copies \cite{vw} and to construct blockspins for the derivation of 
effective theories \cite{hh}.

The eigenvectors are non-local in the sense that they depend on all link variables.
In order to maintain the transfer matrix formalism the $f(x)$ have to be local in time.
This is achieved by considering the spatial Laplacian $D_i^2[U_i]$ in \eq{(\ref{lev}), which
then is defined in every timeslice individually and independent of $U_0$. 

The eigenvectors are used to construct an $N\times N$ matrix 
$\Omega(x)\in SU(N)$ following \cite{vw}. Since \eq (\ref{lev}) only determines them up to
a phase, this leaves a remaining freedom in $\Omega(x)$.
In the case of $SU(2)$, all eigenvalues are two-fold degenerate due to charge conjugation,
and the two vectors to the lowest eigenvalue are combined into $\Omega$,
which then is determined up to a global $SU(2)$ rotation $h$.
For $SU(3)$ there is no degeneracy of the eigenvalues in general. In this case 
one solves for the three lowest eigenvectors to construct the  matrix $\Omega$,
which is then determined 
up to a factor $h={\rm diag}(\exp(\ii \omega_1),\exp(\ii \omega_2),\exp(\ii \omega_3)),
\sum_i\omega_i=0$.
This may be summarized by the transformation behaviour 
\be \label{otrafo}
\Omega^g(x)=g(x)\Omega(x)h^\dag (t), 
\ee
where $h(t)$ is free and may be different in every timeslice.

We can now define composite link and gluon fields
\ba \label{cl}
V_\mu(x)&=&\Omega^\dag(x)U_\mu(x)\Omega(x+\hat{\mu}),\\
A_\mu(x)&=&\frac{\ii}{2g}\left[\left(V_\mu(x)-V^\dag_\mu(x)\right)
-\frac{1}{N}\tr\left(V_\mu(x)-V^\dag_\mu(x)\right)\right],\nn
\ea
both transforming as
$O_i^g(x)=h(t)O_i(x)h^\dag(t)$, whereas $V_0^g(x)=h(t)V_0(x)h^\dag(t+1)$.
Hence the $A_i$ are gauge invariant under spatial transformations $g(\bfx)$, 
but transform under time-dependent rotations corresponding to the residual symmetry
of the spatial Laplacian.

With $V_0(\bfx;t_1,t_2)$ denoting the temporal Wilson line
connecting the  
sites $(\bfx,t_1),(\bfx,t_2)$,
we can now construct the manifestly gauge invariant operator
\be \label{gluestring}
O[U]=\tr \left[A_i(\bfx,0)V_0(\bfy;0,t)
A_i(\bfx,t)V_0^\dag(\bfz;0,t)\right].
\ee
It represents a correlator in $t$
of the composite field $A_i$, where Wilson lines are inserted
to ensure full gauge invariance. Note that these may be placed at any $\bfy,\bfz$.

In order to obtain a spectral decomposition of $\langle O[U] \rangle$ it
has to be converted into a trace over quantum mechanical states. 
In a Hilbert space formulation \cite{ks,cre,pos} a
spatial sublattice $L^d$ at a fixed time is considered, with 
link variables $U(\bfx,i)$.
The wave functions form a Hilbert space ${\h}_0$ of all complex,
square integrable functions
$\psi[U]$ defined on the gauge group $G$:
${\h}_0=[L^2(G)]^{d L^d}$.
Wave functions of physical states are gauge invariant, $\psi[U^g]=\psi[U]$,
forming a subspace $\h\subset$${\cal H}_0$.
Any wave function $\psi[U]\in {\h}_0$ can be projected on the physical subspace by means
of the projection operator $\hp$,
\be
(\hp \psi)[U]=\int \prod_{\bfx} dg(\bfx)\;\psi[U^g].
\ee
The dynamics is introduced
by means of the transfer matrix $\htm$, which translates
wave functions by one lattice spacing in time. It is an integral operator
\ba
\psi[U_{t+1}]&=&\left(\htm \psi\right)[U_t]\nn\\
&=&\int\prod_{\bfx,i}dU_t(\bfx,i) \;K[U_{t+1},U_t]\psi[U_t]\;,
\ea
with kernel
\be \label{tm2}
K[U,U']=\int\prod_{\bfx}dW(\bfx) \exp -S_t[U,W,U'].
\ee
Here $S_t$ is the action of two neighbouring timeslices, and the field $W(\bfx)$ is to be
identified with the timelike links \cite{pos}.
Using the gauge invariance of $S_t$, one shows that
\be
\htm=\hp\htm_0,
\ee
where $\htm_0$ is an integral operator with kernel
$K_0[U,U']=\exp-S_t[U,1,U']$,
which is invariant under time independent gauge transformations $g(\bfx)$.
$\htm_0$ is a bounded, self-adjoint operator
with a strictly positive spectrum \cite{pos},
thus allowing to define a Hamiltonian \cite{pos,cre} $\hat{H}_0=- 1/a\ln \htm_0$,
acting on the Hilbert space ${\cal H}_0$. Through the projection $\htm$
is defined on the gauge invariant subspace ${\cal H}$, on which there 
is a corresponding Hamiltonian
$\hat{H}$. Generally, the spectrum of $\hat{H}$ consists of the physical
particle states of the theory, which couple to local gauge invariant operators.
In addition to these states, $\hat{H}_0$ contains also the spectrum of
gauge field excitations in the presence of static sources, such as the static potential.
The corresponding pure gauge wave functions in one timeslice 
have non-trivial transformation behaviour.

We now define the multiplication operators
$(\hat{\Omega}_{\alpha\beta}\psi)[U]=\Omega_{\alpha\beta}[U]\psi[U]$ and
$(\hat{V}(\bfx,i)\psi)[U]=V(\bfx,i)\psi[U]$, so that
$\hat{A}(\bfx,i)$
is a multiplication operator as well. 

Since $O[U]$ is manifestly gauge invariant, its expectation value is the same
when evaluated in temporal gauge with a fixing function
$F_{0}[U]=\prod_x\delta[U_0(x),1]$.
Writing the quantum mechanical trace over a complete set of states on the space
$\h$ as
$
\trq{\op}=\sum_n\langle n|\op| n \rangle
$,
one finds
\ba 
\langle O[U]\rangle & = & \langle O[U]F_0[U] \rangle \nn\\ 
&\hspace*{-1.5cm}=&\hspace*{-0.5cm} Z^{-1} \trq\left\{\htm^{N_t-t}
\left(\hat{\Omega}(\bfx)\hat{A}(\bfx,i)
\hat{\Omega}^\dag(\bfx)\right)_{\alpha\beta} \right.\;\nn\\ 
&\hspace*{5mm}\times& \left.\hspace*{8mm}\htm_0^t \;
\left(\hat{\Omega}(\bfx)\hat{A}(\bfx,i)\hat{\Omega}^\dag(\bfx)\right)_{\beta\alpha}\right\},
\ea
which has the spectral decomposition
\ba \label{stringen}
\lim_{N_t\rightarrow \infty} \langle O[U] \rangle &= &\sum_n
|\langle 0|\left(\hat{\Omega}(\bfx)\hat{A}(\bfx,i)\hat{\Omega}^\dag(\bfx)\right)_{\alpha\beta}
|n^0\rangle |^2 \nn\\
& &\times \ex^{ -(E_n-E_0)t}\;.
\ea
Here $\{|n^0\rangle\}$ is a complete set of eigenstates of $\htm_0$ and 
the matrix elements are assumed to be non-zero.

On the other hand, one may also 
employ a fixing function 
$F_{L0}[U]=\prod_x\delta[\Omega(x)U_0(x)\Omega(x+\hat{0}),1]$, where now the composite 
$V_0$ are brought to temporal gauge.
This leads to the expression
\ba \label{L0ham}
\langle O[U]\rangle & = & \langle O[U]F_{L0}[U] \rangle \nn\\ &= &
 Z^{-1}\trq\left\{\htm^{N_t-t}\hat{A}(\bfx,i)_{\alpha\beta}\;
\htm_{L0}^t \;
\hat{A}(\bfx,i)_{\beta\alpha}\right\},
\ea
where we have defined a modified transfer matrix $\htm_{L0}$ with 
kernel
\be
K_{L0}[U,U']=\ex^{-S_t[U,\Omega^\dag\Omega',U']}=\ex^{-S_t[U^{\Omega^\dag},1,
U^{'{\Omega'}^\dag}]},
\ee
and the notation $\Omega'=\Omega[U']$.
According to the last equation the integral kernels $K_0,K_{L0}$ are simply related
by 
\be
K_{L0}[U,U']=K_0[U^{\Omega^\dag}, U^{{'\Omega'}^\dag}].
\ee
One easily verifies that $K_{L0}$ shares most properties of $K_0$. 
In particular it is real, symmetric, $K_{L0}[U,U']=K_{L0}[U',U]$,
and square integrable, so that $\htm_{L0}$ is a self-adjoint, compact operator. 
It is gauge invariant under time independent tranformations $g(\bfx)$.
Moreover, the positivity proof for $\htm_0$ in \cite{pos} goes 
through unchanged for $\htm_{L0}$.
Beyond the functional form of $K_0$, it only uses the fact that the product
$U^{\dag}(\bfx,i)U'(\bfx,i)$ is in a fundamental representation of $SU(N)$,
which is also true 
for $U^{\Omega^\dag}(\bfx,i), U^{'{\Omega'}^\dag}(\bfx,i)$.
Thus $\htm_{L0}$ has a strictly positive spectrum.
Finally, the transfer matrix $\htm$ acting on $\h$
is again obtained by projection, $ \htm=\hat{P}\htm_{L0}$.

There then exists a complete set 
$\{|n^{L0}\rangle\}$ 
of eigenstates of $\htm_{L0}$ to obtain the spectral decomposition of \eq (\ref{L0ham}),
\be \label{finalen}
\lim_{N_t\rightarrow \infty} \langle O[U] \rangle = \sum_{n}
|\langle 0|\hat{A}_{\alpha\beta}(\bfx)
|n^{L0}\rangle |^2 \ex^{ -(E'_n-E_0)t}\;.
\ee
\eqs (\ref{stringen}),(\ref{finalen}) are two spectral representations of the same
expectation value $\langle O[U]\rangle$ and of the form
\be \label{series}
\sum_n a_n \ex^{-E_nt}
= \sum_{n}
b_n \ex^{ -E'_nt}\:,
\ee
with $a_n,b_n>0$, for every $t$. In the limit $t\rightarrow \infty$ only the $n=1$ terms 
are retained.
Expanding the exponentials one obtains in this limit
\be 
\sum_m \frac{(-t)^m}{m!} a_1 (E_1)^m
=\sum_m \frac{(-t)^m}{m!}b_1 (E'_1)^m.
\ee 
Both sides of this equation represent a power series in $t$ of the same analytic function of $t$,
hence the coefficients of $t^m$ are the same for every $m$.
It follows from the $m=0$ term that $a_1=b_1$, while the first derivative of the equation 
with respect to $t$ yields 
$a_1E_1=b_1E'_1$, and hence $E_1=E'_1$. Going back to \eq (\ref{series}) and 
subtracting the $n=1$ term on both sides, the procedure may be repeated to establish
$a_n=b_n$ and $E_n=E'_n$ term by term.
We thus conclude that $\htm_{L0}$ has the same spectrum as $\htm_0$, and hence
the constructed correlator falls off exponentially with eigenvalues of the Hamiltonian
$\hat{H}_0$.

The operator \eq (\ref{gluestring}) involves temporal Wilson lines
which may be interpreted as propagators of static sources, and the field energies
extracted from this correlator are those of a gluon field in the presence
of sources.
As the continuum limit is approached,
the energies will thus diverge because of the well-known divergent self-energy
contributions of the temporal Wilson lines, \fig \ref{div}. 
In order to retain a finite continuum limit, the operator
has to be modified such that no divergent mass renormalization 
is present.

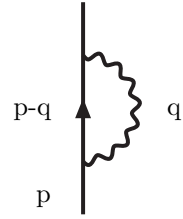
\begin{figure}[tbh]

\vspace*{-0.9cm}

\begin{center}

\begin{picture}(150,120)(0,0)

\SetWidth{1.5}
\ArrowLine(62,20)(62,100)
\PhotonArc(62,60)(20,-90,90){1.5}{8}

\Text(50,26)[r]{p}
\Text(50,59)[r]{p-q}
\Text(94,59)[l]{q}



\end{picture}
\vspace*{-0.4cm}
\end{center}
\caption[]{\label{div}\it
The Wilson line self-energy. \eq (\ref{tav}) enforces $p=p-q=0$.}

\vspace*{0.5cm}

\end{figure}

This can be achieved by observing that the transformation behaviour of $V_0(\bfx,t)$ in \eq
(\ref{cl}) is independent of the spatial coordinates. In the construction of the operator
\eq (\ref{gluestring}), instead of $V_0$ we may then
use its timeslice average
\be \label{tav}
\tilde{V_0}(t)=\sum_{\bfx}V_0(\bfx,t)/||\sum_{\bfx}V_0(\bfx,t)||,
\ee
which has been projected back into the group. 
The timeslice average corresponds to the Fourier transform of $V_0$ with zero momentum. 
If this is done
in every timeslice, the sources represented by the $\tilde{V}_0$ cannot emit 
a gluon at one $t$ and reabsorb it at some later $t$ as in Fig.\ref{div}.
Hence, the mass renormalization of the static source is switched off, and the sources
remain classical external fields.
The presence of fields $\tilde{V}_0$ 
then merely accounts for the transformation behaviour, but has
no effect on the gauge field energies
measured by the modified operator
\be \label{ofinal}
O[U]=\tr \left[A_i(\bfx,0)\tilde{V}_0(0,t)
A_i(\bfx,t)\tilde{V}_0^\dag(0,t)\right],
\ee
which has a spectral decomposition as in \eq (\ref{finalen}).
The energies extracted from the expectation values of 
\eqs (\ref{gluestring}) and (\ref{ofinal}) should then differ 
by a cut-off dependent shift due to the selfenergy contribution \fig \ref{div}.

One may now ask to what extent these results depend on the
particular choice of $\Omega[U]$.
Clearly, any $\Omega\in SU(N)$ local in time and
transforming as in \eq (\ref{otrafo})
permits construction of the gauge invariant observable \eq (\ref{ofinal}). From the
spectral representation it follows that all such observables fall off with the same
spectrum, while $\Omega$ only enters the matrix elements representing
the overlap of the operator with the eigenstates.

The construction of the composite link variable \eq (\ref{cl})
may also be viewed
as fixing Laplacian gauge on each timeslice \cite{vw}.
It is crucial that $\Omega$ depends only on spatial links
to preserve the transfer matrix. In the language of gauge fixing this means
that the gauge is incomplete, with
a global factor $h$ remaining unfixed between time-slices.
It can be completed by imposing
the further condition $\tilde{V}_0(t)=1$, thus fixing $h(t)$.
In this particular gauge the operator
\eq (\ref{ofinal}) reduces to a gauge fixed gluon propagator,
falling off exponentially with eigenvalues of the transfer matrix.
Since the spectrum is unaffected by the particular construction of $\Omega$, the statement
holds for all gauges employing a unique $\Omega[U]$
local in time.
For example, fixing the Coulomb gauge by the standard minimization of
$R[U]=\sum_{x,i}[1-1/N\tr (U^{\Omega^\dag}_i)]$ in every timeslice produces an $\Omega$
with the desired properties and a residual freedom $h(t)\in SU(N)$.
(Of course, this gauge condition has the problem that it does not determine
$\Omega$ uniquely \cite{rev}). On the other hand, Landau gauge is non-local in time and
no positive transfer matrix is defined.

Fourier transforming the spectral representation \eq (\ref{finalen}), 
one derives the K\"allen-Lehmann representation in the limits $N_t\rightarrow \infty,
L^d\rightarrow \infty$,
\ba
\langle O[U] \rangle &=&\int_0^\infty dE\int_{-\pi}^\pi\frac{d^dp}{(2\pi)^d}
\ex^{-Et+\ii\bfp\bfx}\rho(E,\bfp),\nn\\
\rho(E,\bfp)&=&\frac{Z(\bfp)}{2\omega(\bfp)}\delta(E-\omega(\bfp))+\bar{\rho}(E,\bfp).
\ea
Hence the propagator in momentum space has a gauge invariant
pole 
defining a parton mass,
$m=\omega({\bf 0})=E_1-E_0$, with a residue
$\sqrt{Z(\bfp)}=\langle 0|\hat{A}({\bf 0})|\bfp\rangle$,
while $\bar{\rho}(E,\bfp)$ contains all higher states.

It should be stressed here
that this does {\it not} imply an asymptotically free colour charged state
in the spectrum of the theory. All asymptotic one particle states
satisfy Gauss' law without static charges, they
are eigenstates of the projected transfer matrix $\htm$ acting on the Hilbert space
$\h$ of gauge invariant functions.
By contrast, the additional eigenstates $|n^{L0}\rangle\in {\cal H}_0$ 
satisfy Gauss'
law in the presence of static sources and are eigenstates of the transfer matrix $\htm_0$.

All these considerations are easily extended to matter fields. In particular one may 
carry out the same analyis for the gauge invariant quark correlator
\be
 \left\langle\tr\left(\bar{\Psi}(\bfx,0)\Omega(\bfx,0)\tilde{V}_0(0,t)
\Omega^\dag(\bfx,t)\Psi(\bfx,t)\right)\right\rangle. 
\ee
One would expect the mass extracted from its exponential decay to coincide with
the renormalization group invariant quark mass computed in non-perturbative renormalization
schemes \cite{npr}.

In summary, it has been shown that quark and gluon propagators exhibit an exponential
decay governed by eigenvalues of the Hamiltonian in all gauges that are local in time,
implying a corresponding gauge invariant
and non-perturbative singularity structure in momentum space. 
In particular, the first excitation energy over the vacuum, 
viz. the pole closest to the origin, may serve
as non-perturbative definition of a parton mass. 
This promotes propagator based definitions of Debye and magnetic screening in
the QCD plasma to a non-perturbative level, and 
opens similar questions for the zero temperature case.
The analysis may be extended to 
other Green functions and should reveal valuable information about the colour 
dynamics on any distance scale. Examples together with detailed numerical experiments
in three dimensions will be reported elsewhere.

I thank O.B\"ar, P.~de Forcrand and U.-J.~Wiese for numerous discussions, 
and P.~van Baal for useful
comments on the manuscript.

}
\end{multicols}
\end{document}